\documentclass[12pt]{article}
\usepackage{hyperref}
\hypersetup{
colorlinks=true,
linkcolor=black,
citecolor=black,
urlcolor=cyan,
filecolor=black,
pdfborder={0 0 0},
}

\usepackage{amssymb}
\usepackage{amsfonts}
\usepackage{mathrsfs}
\usepackage{latexsym}
\usepackage{amsmath}
\usepackage{graphics}
\usepackage{graphicx}
\usepackage{sectsty}
\usepackage{setspace}
\usepackage{multirow}
\sectionfont{\normalsize\bf}
\subsectionfont{\rm\normalsize}

\usepackage{tikz}
\usetikzlibrary{matrix,arrows,decorations.pathmorphing}

\DeclareMathOperator{\im}{im}

\usetikzlibrary{positioning}

\def\be{\begin{equation}}
\def\ee{\end{equation}}
\def\bd{\left|\begin{matrix}}
\def\ed{\end{matrix}\right|}
\input amssym.def
\input amssym
\def\Cop{{\Bbb C}}
\def\Pop{{\Bbb P}}
\def\M{{\cal M}}
\def\R{{\cal R}}
\def\S{{\cal S}}
\def\T{{\cal T}}
\def\J{{\cal J}}
\def\I{{\cal I}}
\def\K{{\cal K}}
\def\U{{\cal U}}
\def\h{{\boldsymbol h}}
\def\half{{\scriptstyle {1\over 2}}}
\def\bu{\bullet}
\def\hsk{\hskip-8pt}\def\hs{\hskip2pt}
\def\hska{\hskip-4pt}\def\hsa{\hskip6pt}
\numberwithin{equation}{section}
\begin{document}

\thispagestyle{empty}
\begin{flushright}
\end{flushright}
\baselineskip=16pt
\vspace{.5in}
{%\Large
\begin{center}
{\bf General Solution of the Scattering Equations}
\end{center}}
\vskip 1.1cm
\begin{center}
{Louise Dolan}
\vskip5pt

\centerline{\em Department of Physics}
\centerline{\em University of North Carolina, Chapel Hill, NC 27599} 
\bigskip
\bigskip        
{Peter Goddard}
\vskip5pt

\centerline{\em School of Natural Sciences, Institute for Advanced Study}
\centerline{\em Princeton, NJ 08540, USA}
\bigskip
\bigskip
\bigskip
\bigskip
\end{center}

\abstract{\noindent 
The scattering equations, originally introduced by Fairlie and Roberts in 1972 and more recently shown by Cachazo, He and Yuan to provide a kinematic basis for describing tree amplitudes for massless particles in arbitrary space-time dimension, have been reformulated in polynomial form. The scattering equations for $N$ particles are equivalent to $N-3$  polynomial equations  $ h_m=0$, $1\leq m\leq N-3$,  in $N-3$ variables, where $h_m$ has degree $m$ and is linear in the individual variables. Facilitated by this linearity, elimination theory is used to construct a single variable polynomial equation,  $\Delta_N=0$, of degree $(N-3)!$ determining the solutions. 
$\Delta_N$ is the sparse resultant of the system of polynomial scattering equations and  it can be identified as the hyperdeterminant of a multidimensional matrix of border format within the terminology of 
 Gel'fand, Kapranov and Zelevinsky. Macaulay's Unmixedness Theorem is used to show that the polynomials of the scattering equations constitute a regular sequence, enabling the Hilbert series of the variety determined by the  scattering equations to be calculated, independently showing that they  have  $(N-3)!$ solutions.

  }
\bigskip

%\vskip120pt email: ldolan@physics.unc.edu
\setlength{\parindent}{0pt}
\setlength{\parskip}{6pt}

\setstretch{1.05}
\vfill\eject
\vskip50pt
%%%%%%%%%%%%%%%%%%%%%%%%%%%%%%%%PAPER%%%%%%%%%%%%%%%
\section{Introduction}
\label{Introduction}

In this paper, we describe the solution of the scattering equations originally introduced by Fairlie and Roberts in 1972 \cite{FR}  in a search for new dual models (string theories) without tachyons and subsequently found again by Gross and Mende while studying  the high-energy behavior of string theory \cite{GM}. Recently they have been rediscovered by Cachazo, He and Yuan  (CHY)  \cite{CHY0}-\cite{CHY3}, who have shown that they provide a basis for describing the kinematics of massless particles, by proposing remarkable formulae for their tree amplitudes, which have been proved for scalar field theory and gauge theory \cite{DG1}.

We consider $N$ massless particles, labelled by $a\in A=\{1,2,\ldots,N\}$,  with momenta, $k_a, a\in A,$ with $k^2_a=0$ and $\sum_{a\in A}k_a=0$, and
introduce a variable $z_a\in\Cop$ for each $a\in A$. Then the {\it scattering equations} are the $N$ equations $f_a(z,k)=0, a\in A,$ where
\be
  f_a(z,k)=\sum_{b\in A\atop b\ne a}{k_a\cdot k_b\over z_a-z_b}.\label{SE}
\ee
This system of $N$ equations is M\"obius invariant system, and consequently satisfy three relations,
\be 
\sum_{a\in A} f_a(z,k)=0;\qquad\sum_{a\in A}z_a f_a(z,k)=0;\qquad\sum_{a\in A}z_a^2 f_a(z,k)=0,\label{ids}
\ee
so that there are only $N-3$ linearly independent equations in the system.

The equations (\ref{ids}) can clearly be written in polynomial form by multiplying by suitable factors of $z_a-z_b$, but it is desirable to obtain 
a set of polynomials whose degrees are as small as possible. The equations typically have $(N-3)!$ solutions \cite{CHY0}, up to M\"obius transformations, and so B\'ezout's theorem,
which states that the number of solutions of a system of $M$ polynomial equations in $M$ unknowns is the product of the degrees of the polynomials, 
suggests that there should be a system of $N-3$, equations of degrees $1,2, \ldots N-3,$ that is equivalent to the scattering equations, up to M\"obius transformations, and this has been shown to be the case \cite{DG2}.

The proposals of CHY  for tree amplitudes in massless theories were expressed as integrals of rational 
functions of the  $z_a$ and the momenta, around poles occurring at the each of the solutions of  the scattering equations. These integrals are hence just sums   
over the solutions and  thus, necessarily, rational functions of the coefficients in the scattering equations.
Therefore, the prescription is fundamentally one of attaching an algebraic expression to the scattering equations, or, equivalently, the zero-dimensional 
variety that they describe; the integral is somewhat symbolic. The objective is to understand the amplitudes in terms of natural algebraic objects attached to this variety. 

A number of authors have discussed the use of techniques of algebraic geometry and commutative algebra to discuss the polynomial scattering equations \cite{HMS}-\cite{CK2}. There are also extensions of the scattering equations to massive particles \cite{DG1},\cite{N}; discussions of world sheet theories and loop amplitudes \cite{MS}-\cite{CT}, leading to one-loop (or torus) scattering equations that can also be put in polynomial form \cite{DG3}; and  discussions of the loop amplitudes by mapping onto the Riemann sphere \cite{GMMT1}-\cite{HY}, but we shall not discuss any of these further here.

The scattering equations are equivalent \cite{DG2} to the $N-3$ polynomial equations $\tilde h_m(z,k)=0, 2\leq m\leq N-2,$ where 
\be
\tilde h_m(z,k)=\sum_{S\subset A\atop |S|=m}k_S^2z_S, \qquad 2\leq m\leq N-2,\label{PSE}
\ee
 the sum is over all $N!/m!(N-m)!$ subsets $S\subset A$ with $m$ elements, and 
\be
k_S=\sum_{b\in S\atop }k_b,\qquad z_S=\prod_{a\in S}z_a, \qquad S\subset A. \label{kS}
\ee
This can be established by an argument closely related to the original motivation of Fairlie and Roberts for introducing the scattering equations. 
They replaced the Virasoro conditions of string theory by the condition 
\be
p(z)^2=\sum_{a,b\in A}{k_a\cdot k_b \over (z-z_a)(z-z_b)}=0,\label{p2}
\ee
where 
\be
p(z)^\mu=\sum_{a\in A}{k_a^\mu \over z-z_a}.\label{p1}
\ee
$p(z)^2$ has no double poles because $k_a^2=0, a\in A$, and, since it vanishes sufficiently fast at infinity, it will vanish everywhere  provided it has no pole at $z=z_a,$ for each $a\in A$, and this condition is just the scattering equations (\ref{SE}). However, the vanishing of $p(z)^2$ is equivalent to the vanishing of the  polynomial of degree $N-2$, 
\be
F(z)=p(z)^2\prod_{a\in A} (z-z_a) =\sum_{S\subset A\atop |S|=2}k^2_S \prod_{b\in\bar S}(z-z_b),\label{defF}
\ee
where $\bar S=\{a\in A:a\notin S\}$. It is straightforward to show that the coefficients of $z^{N-2}$ and $z^{N-3}$ in $F(z)$ vanish, and that 
\be
F(z)=\sum_{m=2}^{N-2}(-1)^mz^{N-m-2}\tilde h_m.
\ee
establishing the equivalence of the polynomial form (\ref{PSE}) to the original scattering equations (\ref{SE}).

We can fix the  M\"obius invariance of the system partially by taking two of the $z$, $z_1$ and $z_N$, say, to $\infty$ and $0$, respectively.
\be
h_m(z,k)=\lim_{z_1\rightarrow\infty}{\tilde h_{m+1}\over z_1}=\sum_{S\subset A'\atop |S|=m}\sigma_Sz_S,\qquad 1\leq m\leq N-3,\label{defh}
\ee
where $A'=\{a\in A: a\ne 1, N\}$, $\sigma_S=k^2_{S_1}$ and $S_1= S\cup\{1\}$. $h_m$ is an homogeneous polynomial of degree $m$ in the variables $z_2, \ldots, z_{N-1},$ linear in each of them separately.  The equations $h_m(z,k)=0$, $1\leq m\leq N-3$, define a (presumably) zero-dimensional projective variety in $\Cop\Pop^{N-3}$, a set consisting typically of 
\be
\prod_{m=1}^{N-3}\deg h_m=(N-3)! 
\ee
points by B\'ezout's Theorem. Writing $u=z_{N-2}, v=z_{N-1}$, we seek to construct a polynomial equation of degree $\delta_N=(N-3)! $ in $u/v$, whose roots determine the solutions of the scattering equations. 

As we shall see, while the degree of the $h_m$ determine the degree of this polynomial and so the number of solutions, it is the linearity of the $h_m$ in each $z_a$ taken separately that facilitates the derivation of the 
degree $\delta_N$ polynomial in a relatively simple explicit form. As we shall review in section \ref{Elimination}, this has already been done for $N\leq 6$ \cite{DG2}. For $N=4$ and $5$, it is immediate: for $N=4$, there is just one linear equation, $h_1=0$, which just relates $u$ and $v$ and so determines their ratio; for $N=5$, the two equations, $h_1=h_2=0$, provide two  relations  between $x, u,v,$ where $x=z_2$, linear in $x$, which, when $x$ is eliminated between them,  provide the desired quadratic equation for $u/v$. 

For $N=6$, writing $x=z_2, y=z_3,$ the equations $h_1=h_2=h_3=0$, although linear in the individual $z_a,$ now involve $xy$ as well as $x, y$, and so the elimination of $x$ and $y$ is more complicated, but can be achieved by a simple application of the elimination theory developed by Sylvester and Cayley in the 19th century \cite{JJS}--\cite{GS}. We supplement $h_1=h_2=h_3=0$  with $xh_1=xh_2=xh_3=0$ to provide six linear relations
between $1, x, y, xy, x^2$ and $x^2y$, and the condition of their consistency is that the  vanishing of the determinant of the matrix of their coefficients provides the desired sextic in $u/v$, which, as we review in section \ref{Elimination}, we can write the form $\Delta_6=0$ \cite{DG2}, where 
\be
\Delta_6=\bd h_1&h_1^y&h_1^x&h_1^{xy}&0&0\cr h_2&h_2^y&h_2^x&h_2^{xy}&0&0\cr h_3&h_3^y&h_3^x&h_3^{xy}&0&0\cr  0&0&h_1&h_1^y&h_1^x&h_1^{xy}\cr 0&0&h_2&h_2^y&h_2^x&h_2^{xy}\cr  0&0&h_3&h_3^y&h_3^x&h_3^{xy}\cr\ed,\label{Delta6}
\ee
with  $h_m^x=\partial_xh_m, h_m^{x}=\partial_x\partial_yh_m,$ {\it etc.} The advantage of this form  is that we can see directly that $\Delta_6=0$ provides a sextic in $u/v$, vanishing when $h_1=h_2=h_3=0$, and so determining the solutions of the scattering equations, without relying on its derivation using elimination theory.

The sextic $\Delta_6$ is the {\it resultant} of $h_1, h_2, h_3$, regarded as polynomials in $x,y$. The resultant is the polynomial of lowest degree, with integral coefficients, whose variables are the coefficients of $h_1, h_2, h_3$, (which coefficients are themselves in this case homogeneous polynomials in $u,v$, at most linear in either) and which vanishes  when the corresponding three equations have a common solution for $x,y.$ As we discuss in section \ref{Resultant}, from elimination theory, we know that  the resultant is irreducible (that is has no factors other than $\pm 1$) and is unique up to sign.

We shall see in section \ref{Elimination7} that  the discussion for $N=6$ extends  to $N=7$, where we seek a polynomial equation of degree $\delta_7=24$, ideally obtained as the determinantal condition for the consistency of 24 independent linear relations between 24 monomials in $x,y,z,$ where $x=z_2, y=z_3, z=z_4, $ the equations being obtained by multiplying the polynomial scattering equations $h_m, 1\leq m\leq 4,$ by suitable monomials. 

The general approach of elimination theory, as outlined in Cayley's classic article of 1848, which prefigured much of the development of commutative algebra \cite{GKZ}, is to obtain new equations by multiplying the polynomial equations to be solved, treating these equations as linear relations between monomials, until the equations are overdetermined and so yield a consistency condition. It may be that the polynomial equations are not linearly independent and Cayley describes how to take account of the corresponding linear relations between the equations, and, potentially linear relations between these relations and so on. Of course, it is best if such complications can be avoided and fortunately we are able to do this for the polynomial scattering equations. 

For $N=7$, we use the 24 equations given by 
\be
h_m=yh_m=xh_m=xyh_m=x^2h_m=x^2yh_m=0,\qquad  1\leq m\leq 4,\label{PSE7}
\ee
which provide linear relations between the 24 monomials 
\be
x^py^qz^r, \qquad 0\leq p\leq 3, \;0\leq q\leq 2, \;0\leq r\leq 1, 
\ee
compromising 23 variables together with 1. The determinantal condition for consistency of the equations (\ref{PSE7}), which we can show are linearly independent, provides the desired polynomial equation of degree 24 for $u/v$, which we can write in a form like (\ref{Delta6}).

As we show in section \ref{EliminationG}, these results extend to general $N$. Writing $x_a=z_{a+1}, 1\leq a\leq N-4, $
we consider the $(N-3)!$ equations, $\alpha h_m=0, 1\leq m\leq N-3, \alpha\in C_{N-5},$ where $C_M$ denotes the set of monomials, 
\be
C_M=\left\{\prod_{a=1}^{M} x_a^{m_a}: \;  0\leq m_a\leq M-a+1,\; 1\leq a\leq M\right\}.\label{CM}
\ee
These provide linear relations between the $(N-3)!$ monomials $\beta\in C_{N-4},$ including unity. The resulting determinantal consistency condition is the required polynomial, $\Delta_N$, of order $\delta_N$ in $u/v$.  Each element of the determinant can be taken to be a partial derivative of one of the $h_m$ as in (\ref{Delta6}); this is specified in (\ref{MNab}).

Because  $h_m$ is linear in each of the $z_a, 2\leq a\leq N-1$, it is not a general sum of monomials in the $z_a$, and, for this reason it is not the theory of resultants for general multinomials
that is relevant to the solution of the scattering equations but, as we outline in section \ref{Resultant}, the theory of {\it sparse resultants}. Further, the sparse resultant of multilinear equations of the form of the scattering 
equations is given by the hyperdeterminant of a multidimensional array, within the theory developed by  Gel'fand, Kapranov and Zelevinsky \cite{GKZ}. Multilinear systems of this sort occur, {\it e.g.} in discussions of Nash equilibria in game theory (see chapter 6 of \cite{Sturm}); hyperdeterminants have previously occurred in theoretical physics, {\it e.g.} in the context of black hole entropy  \cite{MJD}.

Information about the number of solutions of the scattering equations is encoded in the Hilbert series of the variety they determine. We calculate the Hilbert series by using Macaulay's Unmixedness Theorem \cite{Mac2} to 
show that the polynomials of the scattering equations constitute a regular sequence, and this provides an independent determination that they have  $(N-3)!$ solutions.

Some of our resultants overlap with work of Cardona and Kalousios \cite{CK}, who also obtained the formula for the  polynomial of order $\delta_N$, using an approach based on elimination theory.

\section{Review of the Cases $N=4,5$ and $6$}
\label{Elimination}

As described in section \ref{Introduction}, in order to solve the scattering equations $f_a(z,k)=0, a\in A,$ or, equivalently, $h_m(z,k)=0,
1\leq m\leq N-3$, we seek to eliminate $z_a, 2\leq a\leq N-3$, in favor of $u=z_{N-2}$ and $v=z_{N-1};$  to give a polynomial of order $(N-3)!$ in $u/v$.
As before, when appropriate, write $x =z_2$ and $ y=z_3$.

\subsection{\it Solutions for $N=4$ and $N=5$.}  
For $N=4$, we have one simple linear equation determining $u/v$,
\be
h_1=\sigma_2u+\sigma_3v=0,\quad u/v=-\sigma_3/\sigma_2=-k_1\cdot k_3/k_1\cdot k_2.
\ee

For $N=5$, eliminating $x$ between the equations
\begin{align}
h_1&=\sigma_2x+\sigma_3u+\sigma_4v=0\cr
h_2&=\sigma_{23}xu+\sigma_{24}xv+\sigma_{34}uv=0
\end{align}
yields a quadratic equation for $u/v$ given by the vanishing of the determinant
\be
\bd \sigma_3u+\sigma_4v&\sigma_2\cr
\sigma_{34}uv&\sigma_{23}u+\sigma_{24}v\cr\ed
=\bd h_1&h_1^{x}\cr h_2&h_2^{x}\ed,
\ee
where $h_m^x=\partial_xh_m,$ as we see by adding $x$ times the second column to the first. Now, we can 
see directly that $\Delta_5=0,$ where 
\be
\Delta_5
=\bd h_1&h_1^{x}\cr h_2&h_2^{x}\ed,\label{Delta5}
\ee
gives the required quadratic in $u/v$ because (i) it vanishes when $h_1=h_2=0$;
(ii) it is a homogeneous quadratic in $(x,u,v)$;  (iii) it is independent of $x$. To see that $\Delta_5$ is 
independent of $x$, note that
\be
\partial_x\Delta_5=\bd h_1^x&h_1^{x}\cr h_2^x&h_2^{x}\ed+\bd h_1&h_1^{xx}\cr h_2&h_2^{xx}\ed=0,
\ee
since $h_m^{xx}=\partial_xh^x_m=0,$ because $h_m$ is linear in $x$. To see that $\Delta_5$ is quadratic,
note that the entry in row $r$ and column $c$ is of degree $r-c+1$; since the sum of the row numbers equals 
the sum of the column numbers for any product contributing to the determinant, $\Delta$ is of degree $2$. Note
that the term in $v^2$ comes from the product $h_1h_2^x$ and its coefficient is $\sigma_4\sigma_{24}$.

Another way, which will be useful when we consider larger values of $N$,  to establish the elementary fact that $\Delta_5$ 
is independent of $x$ is as follows. We are seeking the condition on $u,v$ (and the coefficients in the $h_m$)
that $h_1(x,u,v)=0$ and $h_2(x,u,v)=0$ have a common solution for some value of $x$. For any given value of $x_0$, this is
clearly the same as the condition that $h_1(x_0+\xi,u,v)=0$ and $h_2(x_0+\xi,u,v)=0$ have a common solution for some value of $\xi$.
So this condition must be independent of $x_0$. But
\be
h_m(x_0+\xi,u,v)=h_m(x_0,u,v)+h_m^x(x_0,u,v)\xi,\qquad m=1,2,
\ee
and the condition for these equations to have a common solution for $\xi$ is $\Delta_5=0$ evaluated at $(x,u,v)=(x_0,u,v)$. So this 
condition is independent of $x_0$, which implies that $\Delta_5$, as defined by (\ref{Delta5}), is independent of $x$.

Since $h_1,h_2$ and $h_2^x$ are linear in $v$ and $h_1^x$ is constant, the leading power of $v$  in $\Delta_5$,  $v^2$, comes from the product $h_1h_2^x$ and its
coefficient is 
\be
\sigma_4\sigma_{24}. \label{v2}
\ee
The corresponding result for $N=4$ is that the coefficient of $v$ in $\Delta_4=h_1$ is $\sigma_3$. These leading coefficients are
nonzero if all the  Mandelstam variables $k_S^2, S\subset A,$ are nonzero.

\subsection{\it Solution for $N=6$.}  

For $N=6,$ we seek to eliminate  $x$ and $y$ between the equations
\begin{align}
h_1&=\sigma_2x+\sigma_3y+\sigma_4u+\sigma_5v=0,\cr
h_2&=\sigma_{23}xy+\sigma_{24}xu+\sigma_{25}xv+\sigma_{34}yu+\sigma_{35}yv+\sigma_{45}uv=0,\cr
h_3&=\sigma_{234}xyu+\sigma_{235}xyv+\sigma_{245}xuv+\sigma_{345}yuv=0,
\end{align}
to obtain a sextic in $u,v$. We shall see that obtaining a simple formula for the polynomial in $u,v$ depends on 
each $h_m$ being linear in each $z_a$, while the degree of this polynomial depends on the degree of $h_m$ equalling $m$.
We can write 
\be
h_m=a_m+b_my+c_mx+d_mxy,\qquad m=1,2,3,
\ee
where in fact $d_1=a_3=0$. Following the approach of elimination theory, we consider the equations, $h_m=0, 1\leq m\leq 3,$ together with the equations
$xh_m=0, 1\leq m\leq 3,$ which provide six homogenous {\it linear} equations relating the five quantities $y, x, xy, x^2, x^2y,$ the consistency of which leads to the
condition $\Delta_6=0,$ where
\be
\Delta_6=\bd a_1&b_1&c_1&d_1&0&0\cr a_2&b_2&c_2&d_2&0&0\cr  a_3&b_3&c_3&d_3&0&0\cr  0&0&a_1&b_1&c_1&d_1\cr 0&0&a_2&b_2&c_2&d_2\cr  0&0&a_3&b_3&c_3&d_3\cr\ed.\label{Delta6o}
\ee
By row and column operations, we can show that we can rewrite $\Delta_6$ as 
\be
\Delta_6=\bd h_1&h_1^y&h_1^x&h_1^{xy}&0&0\cr h_2&h_2^y&h_2^x&h_2^{xy}&0&0\cr h_3&h_3^y&h_3^x&h_3^{xy}&0&0\cr  0&0&h_1&h_1^y&h_1^x&h_1^{xy}\cr 0&0&h_2&h_2^y&h_2^x&h_2^{xy}\cr  0&0&h_3&h_3^y&h_3^x&h_3^{xy}\cr\ed,\label{Delta6a}
\ee
where $h^{xy}_m=\partial_x\partial_y h_m$. We can establish the condition $\Delta_6=0$ in the form (\ref{Delta6a}), following the approach we used for $N=5$, by considering the condition for
$h_m(x+\xi,y+\eta,u,v)=0, 1\leq m\leq 3,$ to have a common solution for $\xi,\eta,$  given values of $x,y,u,v$, noting that this condition is clearly independent of the values of $x,y$. Since
\be
h_m(x+\xi,y+\eta,u,v)=h_m+h_m^y\eta+h_m^x\xi+h_m^{xy}\xi\eta,\qquad m=1,2,3,
\ee
where $h_m=h_m(x,y,u,v),$ {\it etc.}, giving the vanishing of (\ref{Delta6a}) as the condition. 

\subsection{\it Direct demonstration that $\Delta_6$ is the required sextic.}

Again, we can see directly, without using elimination theory, that $\Delta_6,$ as defined by (\ref{Delta6a}), gives the required sextic in $u/v$ because (i) it vanishes when $h_1=h_2=h_3=0$;
(ii) it is a homogeneous sextic in $(x,y,u,v)$;  (iii) it is independent of $x,y$. 

To see that $\Delta_6$ is a homogeneous sextic,
note again that the entry in row $r$ and column $c$, if non-zero, is of degree $m_r-n_c+1$ in $x,y,u,v$, where $(m_1,m_2,m_3,m_4,m_5,m_6)=(1,2,3,2,3,4)$ and 
$(n_1,n_2,n_3,n_4,n_5,n_6)=(1,2,2,3,3,4)$. Since any product contributing to the determinant involves one element from each row and one from each 
column, the degree of the product is $6$ plus the sum of the $m_r$ less  the sum of $n_c$ , {\it i.e.} it equals 6, showing that $\Delta_6$ is a homogeneous sextic.
Because no element in the representation (\ref{Delta6a}) is more than linear in $v$, the term in $v^6$ must be the product of linear factors, and it is then 
easy to see that it comes only from  the product $h_1h_2^yh_3^{xy}h_1h_2^xh_3^{xy}$ and its coefficient is 
\be\sigma_5^2\sigma_{25}\sigma_{35}\sigma_{235}^2.\label{v6}\ee
It is immediately obvious that $\partial_y\Delta_6=0,$ because the $y$-derivative of any column either vanishes by linearity of the entries in $y$, or equals some other column. To show directly that $\partial_x\Delta_6=0$ in a way that will generalize to $N>6$, first write 
\be
h^\alpha=\left(\begin{matrix} h^\alpha_1\cr h^\alpha_2\cr h^\alpha_3\end{matrix}\right)
\ee
where $\alpha = \varnothing,x,y,$ or $xy,$ with $h^\varnothing_m\equiv h_m$. Then
\be
\Delta_6=\bd h&h^y&h^x&h^{xy}&0&0\cr  0&0&h&h^y&h^x&h^{xy}\cr \ed,
\ee
and, subtracting $x$ times column 6 from column 4, and $x$ times column 5 from column 3, and then
$x$ times column 4 from column 2, and $x$ times column 3 from column 1, 
\be
\Delta_6=\bd h_o&h^y_o&h^x&h^{xy}&0&0\cr  -xh_o&-xh^y_o&h_o&h^y_o&h^x&h^{xy}\cr \ed,
\ee
where $h_o=h-xh^x$ and $h_o^y=h^y-xh^{xy}$;
then adding $x$ times row 1 to row 2, and then, again, subtracting $x$ times column 6 from column 4, and $x$ times column 5 from column 3,
we obtain
\be
\Delta_6=\bd h_o&h^y_o&h^x&h^{xy}&0&0\cr  0&0&h_o&h^y_o&h^x&h^{xy}\cr \ed,
\ee
which is manifestly independent of $x$.

\subsection{\it Determination of $x,y$ in terms of $u,v$ satisfying $\Delta_6=0$.}

Given $u,v$ satisfying $\Delta_6=0$, $x,y$ can be determined from linear relations. Consider
\be
\Gamma_6=\bd h&h^x&h^{xy}\ed;\label{Gamma}
\ee
then $\partial_x\Gamma_6=0$ and
\be
\partial_y\Gamma_6=\bd h^y&h^x&h^{xy}\ed,\qquad \partial_y^2\Gamma_6=0.
\ee
Thus  $\Gamma_6$ is independent of $x$, linear in $y$ and vanishes when $h_m=0, 1\leq m\leq 3,$
and so provides a linear relation to determine $y$, given $u,v$ satisfying $\Delta_6=0$. Similarly, $\bd h&h^y&h^{xy}\ed=0$ provides a linear relation, independent of $y$,
to determine $x$.

We can derive the expression (\ref{Gamma}) for $\Gamma_6$ as follows. Given $u,v$ satisfying $\Delta_6=0$, and any $x,y$, we can find $\xi,\eta$ such that 
 $h(x+\xi,y+\eta,u,v)=0$, implying $\Delta_6=0$.  Then $(1,\eta,\xi,\xi\eta,\xi^2,\xi^2\eta)^T$ is a null vector of the matrix obtained by putting $h=h(x,y,u,v)$ in 
 the matrix corresponding to the right hand side of (\ref{Delta6a}) and 
\begin{align}
\eta={\xi\eta\over\xi}&=-\bd  h_2&h_2^y&h_2^x&0&0\cr h_3&h_3^y&h_3^x&0&0\cr  0&0&h_1&h_1^x&h_1^{xy}\cr 0&0&h_2&h_2^x&h_2^{xy}\cr  0&0&h_3&h_3^x&h_3^{xy}\cr\ed
\bd  h_2&h_2^y&h_2^{xy}&0&0\cr h_3&h_3^y&h_3^{xy}&0&0\cr  0&0&h_1^y&h_1^x&h_1^{xy}\cr 0&0&h_2^y&h_2^x&h_2^{xy}\cr  0&0&h_3^y&h_3^x&h_3^{xy}\cr\ed^{-1}\cr
&=-\bd  h_1&h_1^x&h_1^{xy}\cr h_2&h_2^x&h_2^{xy}\cr  h_3&h_3^x&h_3^{xy}\cr\ed
\bd   h_1^y&h_1^x&h_1^{xy}\cr h_2^y&h_2^x&h_2^{xy}\cr  h_3^y&h_3^x&h_3^{xy}\cr\ed^{-1}.
\end{align}
But, if $x,y$ are such that $h(x,y,u,v)=0$, $\eta=0$ and  $\bd h&h^x&h^{xy}\ed=0$ determines $y$ given $u,v$ satisfying $\Delta_6=0$.

\section{Elimination Theory for $N=7$}
\label{Elimination7}

\subsection{\it Construction of the 24th order polynomial $\Delta_7$ using elimination theory}

Writing $(z_2,z_3,z_4,z_5,z_6)=(x,y,z,u,v)$, we seek to eliminate  $x,y$ and $z$ between the equations
$h_m=0, 1\leq m\leq 4,$ to obtain a homogeneous polynomial of order 24  in $u,v$. 
We can write 
\be
h_m=a_m+b_mz+c_my+d_mx+e_myz+f_mxz+g_mxy+j_mxyz,\quad 1\leq m\leq 4, \label{hm4}
\ee
and the column vector $h=(h_1,h_2,h_3,h_4)^T;$ then $h=0$ provides 4 linear relations between 7 variables $x,y,z,xy,zx,yz,xyz$. Adding the 4 equations
$xh=0$ brings in the additional variables $x^2, x^2y, zx^2, x^2yz$; further adding $yh=0$ brings in the variables $y,xy^2,y^2z,xy^2z$, giving a total of 15 
variables with 12 linear relations, still leaving 3 more variables than relations. Then adding the 4 equations $xyh=0,$ just adds two more variables $x^2y^2, 
x^2y^2z, $ leaving one more variable than linear relation. Adding $x^2h=0$ leaves this balance unchanged with 4 more variables, $x^3, x^3y, zx^3, x^3yz,$
to match the 4 new relations. But, finally, adding the 4 relations $x^2yh=0$ adds just the two variables $x^3y^2, x^3y^2z, $ so giving a total of 23 
variables with 24 linear relations, and so a consistency condition $\Delta_7=0,$ where $\Delta_7$ is the $24$ dimensional determinant
\be
\bd
\hs h&\hsk h^z&\hsk h^y&\hsk h^{yz}&\hsk 0&\hsk 0&\hsk h^{x}&\hsk h^{xz}&\hsk h^{xy}&\hsk h^{xyz}&\hsk 0&\hsk 0&\hsk 0&\hsk 0&\hsk 0&\hsk 0&\hsk 0&\hsk 0&\hsk 0&\hsk 0&\hsk 0&\hsk 0&\hsk 0\hsk &\hsk0\cr 
0&\hsk0&\hsk h&\hsk h^z&\hsk h^y&\hsk h^{yz}&\hsk 0&\hsk 0&\hsk h^{x}&\hsk h^{xz}&\hsk h^{xy}&\hsk h^{xyz}&\hsk 0&\hsk 0&\hsk 0&\hsk 0&\hsk 0&\hsk 0&\hsk 0&\hsk 0&\hsk 0&\hsk 0&\hsk 0\hsk &\hsk0\cr 
0&\hsk 0&\hsk 0&\hsk 0&\hsk 0&\hsk 0&\hsk h&\hsk h^z&\hsk h^y&\hsk h^{yz}&\hsk 0&\hsk 0&\hsk h^{x}&\hsk h^{xz}&\hsk h^{xy}&\hsk h^{xyz}&\hsk 0&\hsk 0&\hsk 0&\hsk 0&\hsk 0&\hsk 0&\hsk 0 &\hsk0\cr 
0&\hsk 0&\hsk 0&\hsk 0&\hsk 0&\hsk 0&\hsk 0&\hsk 0&\hsk h&\hsk h^z&\hsk h^y&\hsk h^{yz}&\hsk 0&\hsk 0&\hsk h^{x}&\hsk h^{xz}&\hsk h^{xy}&\hsk h^{xyz}&\hsk 0&\hsk 0&\hsk 0&\hsk 0&\hsk 0 &\hsk0\cr 
0&\hsk 0&\hsk 0&\hsk 0&\hsk 0&\hsk 0&\hsk 0&\hsk 0&\hsk 0&\hsk 0&\hsk 0&\hsk 0&\hsk h&\hsk h^z&\hsk h^y&\hsk h^{yz}&\hsk 0&\hsk 0&\hsk h^{x}&\hsk h^{xz}&\hsk h^{xy}&\hsk h^{xyz}&\hsk 0 &\hsk0\cr   
0&\hsk 0&\hsk 0&\hsk 0&\hsk 0&\hsk 0&\hsk 0&\hsk 0&\hsk 0&\hsk 0&\hsk 0&\hsk 0&\hsk 0&\hsk 0&\hsk h&\hsk h^z&\hsk h^y&\hsk h^{yz}&\hsk 0 &\hsk0&\hsk h^{x}&\hsk h^{xz}&\hsk h^{xy}&\hsk h^{xyz}\cr  \ed,\label{Delta7}
\ee
where $h^{xyz}_m=\partial_x\partial_y\partial_zh_m.$ In principle, $h_m$ and its derivatives should be evaluated at $x=y=z=0$, but we need not impose this condition because, as in the $N=5,6$ cases, for general $x,y,z,$ 
the vanishing of (\ref{Delta7}) provides the condition that $h(x+x',y+y',z+z',u,v)=0$ has solutions for some values of $x',y',z'$, and this is independent of $x,y,z,$ and there is no need to impose the conditions $x=y=z=0$.

In (\ref{Delta7}), the rows correspond to $h,yh,xh,xyh,x^2h$ and $x^2yh,$ respectively, and the columns to 
$ 1, z, y, {yz}, y^2, {y^2z},{x}, {xz},{xy}, {xyz}, {xy^2}, {xy^2z}, {x^2}, {x^2z},{x^2y}, {x^2yz},{x^2y^2}, {x^2y^2z}, {x^3}, {x^3z},{x^3y}, {x^3yz},{x^3y^2}$ and ${x^3y^2z},$ respectively.
Defining $C_N$ as in (\ref{CM}), we
can label the rows of (\ref{Delta7}) by $(\alpha, i), \alpha\in C_2, 1\leq i\leq 4,$ and its columns by $ \beta\in C_3.$ [Note $|C_2|=6, |C_3|=24$.] Then, if 
$B_3=\{x^my^nz^p:  m, n, p=0, 1\}$, the set of monomials in $x,y,z$ in (\ref{hm4}), the elements of the matrix, $M^{(7)}$ corresponding to (\ref{Delta7}),
can be specified by 
\begin{alignat}{3}
M^{(7)}_{\alpha i,\beta}&= h_i^\gamma\equiv\partial_\gamma h_i,\qquad&&\hbox{if } \beta=\alpha\gamma, \;\gamma\in B_3,\\
&= 0,\qquad&&\hbox{if } \beta\notin\alpha  B_3,\
\end{alignat}
where $\partial_\gamma=\partial_x^m\partial_y^n\partial_z^p$ for $\gamma=x^my^nz^p$.  Then
\be
\deg M^{(7)}_{\alpha i,\beta}=i+\deg \alpha-\deg\beta,
\ee
with $M^{(7)}_{\alpha i,\beta}=0$ if $\deg\beta-\deg \alpha=\deg \gamma>i$.
It follows that $\Delta_7=\det M^{(7)}$ is homogeneous of degree 
\be
\sum_{i=1}^4\sum_{\alpha\in C_2}(i+\deg \alpha)-\sum_{\beta\in C_3}\deg\beta=6\times 10 +4\times 9-72=24
\ee
in $x,y,z,u,v$. Again, since no element of $M^{(7)}$ is more than linear in $v$, the term in $v^{24}$ must come from the product of linear factors. 
It is straightforward to see that there is only one such product contributing to $\Delta_7$, namely, written in order of rows,
\be
h_1h_2^zh_3^{yz}h_4^{xyz}h_1h_2^yh_3^{yz}h_4^{xyz}h_1h_2^zh_3^{zx}h_4^{xyz}h_1h_2^yh_3^{xy}h_4^{xyz}h_1h_2^xh_3^{zx}h_4^{xyz}h_1h_2^xh_3^{xy}h_4^{xyz},
\ee
and so the coefficient of $v^{24}$ in $\Delta_7$ is
\be
\sigma_6^6\sigma_{26}^2\sigma_{36}^2\sigma_{46}^2\sigma_{236}^2\sigma_{346}^2\sigma_{246}^2\sigma_{2346}^6.\label{v24}
\ee
 [For further details, see section \ref{Leading}.] 
In particular, it follows that $M^{(7)}$ is non-singular, or, equivalently, the equations $\alpha h_m=0, \alpha\in C_2, 1\leq m\leq 4,$
are linearly independent, and that thus $\Delta_7=0$ is the desired polynomial of order 24 in $u/v$.

\subsection{\it Direct demonstration that $\Delta_7$ is the required polynomial of degree 24.}
\label{directdem7}

So we can see directly from (\ref{Delta7}) that $\Delta_7$ vanishes when $h=0$ and is a homogenous polynomial of degree 24 in $x,y,z,u,v.$
As in the $N=5,6, $ we can show directly that $\Delta_7$ is independent of $x,y,z.$ For this purpose, it is notationally convenient to write $\h$ 
for the $4 \times 2$ matrix $(h,h^z)$, and $\h^\gamma=\partial_\gamma\h, \gamma= x,y$ or $xy$.
\be
\Delta_7=\bd
\hsa \h&\hska \h^y&\hska 0&\hska \h^{x}&\hska \h^{xy}&\hska 0&\hska 0&\hska 0&\hska 0&\hska 0&\hska 0&\hska 0 \cr 
0&\hska \h&\hska \h^y&\hska 0&\hska \h^{x}&\hska \h^{xy}&\hska 0&\hska 0&\hska 0&\hska 0&\hska 0&\hska 0\cr 
0&\hska 0&\hska 0&\hska \h&\hska \h^y&\hska 0&\hska \h^{x}&\hska \h^{xy}&\hska 0&\hska 0&\hska 0&\hska 0 \cr 
0&\hska 0&\hska 0&\hska 0&\hska \h&\hska \h^y&\hska 0&\hska \h^{x}&\hska \h^{xy}&\hska 0&\hska 0&\hska 0 \cr 
0&\hska 0&\hska 0&\hska 0&\hska 0&\hska 0&\hska \h&\hska \h^{y}&\hska 0&\hska \h^{x}&\hska \h^{xy}&\hska 0 \cr   
0&\hska 0&\hska 0&\hska 0&\hska 0&\hska 0&\hska 0&\hska \h&\hska \h^y&\hska 0&\hska \h^{x} &\hska \h^{xy}\cr \ed,\qquad
\h=\left(\begin{matrix}
h_1&h_1^z\cr h_2&h_2^z\cr h_3&h_3^z\cr h_4&h_4^z\cr 
\end{matrix}\right).
\label{Delta7a}
\ee
It follows immediately that $\partial_z\Delta_7=0$. We can show $\partial_y\Delta_7=\partial_x\Delta_7=0$ using the same method that we used to show $\partial_x\Delta_6=0$. To 
show $\Delta_7$ is independent of $y$, first take $y$ times 
the $n$-th column from the $(n-1)$-th, for $n=3,6,9,12$, and then do the same for $n=2,5,8,11$, leading to
\be
\Delta_7=\bd
\hsa \h_\bu&\hska \h^y&\hska 0&\hska \h^{x}_\bu&\hska \h^{xy}&\hska 0&\hska 0&\hska 0&\hska 0&\hska 0&\hska 0&\hska 0 \cr 
-y\h_\bu&\hska \h_\bu&\hska \h^y&\hska -y\h^x_\bu&\hska \h^{x}_\bu&\hska \h^{xy}&\hska 0&\hska 0&\hska 0&\hska 0&\hska 0&\hska 0\cr 
0&\hska 0&\hska 0&\hska \h_\bu&\hska \h^y&\hska 0&\hska \h^{x}_\bu&\hska \h^{xy}&\hska 0&\hska 0&\hska 0&\hska 0 \cr 
0&\hska 0&\hska 0&\hska -y\h_\bu&\hska \h_\bu&\hska \h^y&\hska -y\h^x_\bu&\hska \h^{x}_\bu&\hska \h^{xy}&\hska 0&\hska 0&\hska 0 \cr 
0&\hska 0&\hska 0&\hska 0&\hska 0&\hska 0&\hska \h_\bu&\hska \h^{y}&\hska 0&\hska \h^{x}_\bu&\hska \h^{xy}&\hska 0 \cr   
0&\hska 0&\hska 0&\hska 0&\hska 0&\hska 0&\hska -y\h_\bu&\hska \h_\bu&\hska \h^y&\hska -y\h^x_\bu&\hska \h^{x}_\bu &\hska \h^{xy}\cr \ed,
\label{Delta7b}
\ee
where $\h_\bu=\h - y\h^{y}, \h_\bu^x=\h^x - y\h^{xy}$, which are independent of $y$, by linearity.
Now add $y$ times the $n$-th row to the $(n+1)$-th row for $n=1,3,5$, and then again subtract the $n$-th column from the $(n-1)$-th, for $n=3,6,9,12$, leading to
\be
\Delta_7=\bd
\hsa \h_\bu&\hska \h^y&\hska 0&\hska \h^{x}_\bu&\hska \h^{xy}&\hska 0&\hska 0&\hska 0&\hska 0&\hska 0&\hska 0&\hska 0 \cr 
0&\hska \h_\bu&\hska \h^y&\hska 0&\hska \h^{x}_\bu&\hska \h^{xy}&\hska 0&\hska 0&\hska 0&\hska 0&\hska 0&\hska 0\cr 
0&\hska 0&\hska 0&\hska \h_\bu&\hska \h^y&\hska 0&\hska \h^{x}_\bu&\hska \h^{xy}&\hska 0&\hska 0&\hska 0&\hska 0 \cr 
0&\hska 0&\hska 0&\hska 0&\hska \h_\bu&\hska \h^y&\hska 0&\hska \h^{x}_\bu&\hska \h^{xy}&\hska 0&\hska 0&\hska 0 \cr 
0&\hska 0&\hska 0&\hska 0&\hska 0&\hska 0&\hska \h_\bu&\hska \h^{y}&\hska 0&\hska \h^{x}_\bu&\hska \h^{xy}&\hska 0 \cr   
0&\hska 0&\hska 0&\hska 0&\hska 0&\hska 0&\hska 0&\hska \h_\bu&\hska \h^y&\hska 0&\hska \h^{x}_\bu &\hska \h^{xy}\cr \ed,\label{Delta7c}
\ee
which is manifestly independent of $y$.
A similar, but longer, argument shows that $\partial_x\Delta_7=0$. Thus we have shown directly that $\Delta_7,$ as defined by (\ref{Delta7}), (i) vanishes when $h_1=h_2=h_3=h_4=0$;
(ii)  is a homogeneous 24th order polynomial;  and (iii) is independent of $x,y,z,$  and so provides the required polynomial obtained by eliminating $x,y,z,$ from $h_1=h_2=h_3=h_4=0$.

\section{Elimination Theory for General $N$}
\label{EliminationG}

\subsection{\it Construction of the $(N-3)!$-th degree polynomial $\Delta_N$ using elimination theory}
\label{construction}

For $N\geq 5$, write $x_a=z_{a+1}, 1\leq a\leq N-4, u=z_{N-2}, v=z_{N-1}. $ We seek to eliminate $x_a, 2\leq a\leq N-3,$ between the equations $h_m=0, 1\leq m\leq N-3$. Following the approach adopted 
in the $N=6,7$ cases, we consider the $(N-3)!$ linear relations
\be
h_m\prod_{a=1}^{N-5} x_a^{m_a}=0, \qquad 1\leq m\leq N-3, \quad 0\leq m_a\leq N-4-a,\quad 1\leq a\leq N-5, \label{relns}
\ee
between the variables
\be
\prod_{a=1}^{N-4} x_a^{m_a}=0, \qquad  0\leq m_a\leq N-3-a,\quad 1\leq a\leq N-4,
\ee
that is $(N-3)!-1$ variables other than 1. So, provided that the relations (\ref{relns}) are independent, we can use them to 
eliminate the $x_a, 2\leq a\leq N-3,$ and obtain a polynomial equation for $u/v$.

Let 
\be
C_M=\left\{\prod_{a=1}^{M} x_a^{m_a}: \;  0\leq m_a\leq M-a+1,\; 1\leq a\leq M\right\},\label{defCM}
\ee
\be
R_M=\left\{(\alpha,m): \;  \alpha\in C_{M-1},\; 1\leq m\leq M+1\right\},\label{defRM}
\ee
and
\be
B_M=\left\{\prod_{a=1}^{M} x_a^{m_a}: \;  0\leq m_a\leq 1,\; 1\leq a\leq M\right\}.\label{defBM}
\ee
Note $|C_M|= (M+1)!, |B_M|=2^M$, and
 \be
d_M= \sum_{\alpha\in C_M}\deg\alpha={M(M+1)\over 4}(M+1)!
 \ee
which follows because $d_M$ satisfies the recurrence relation 
\be
d_{M+1}=(M+2) d_M+\half (M+1)(M+2)|C_M|.
\ee

The relations (\ref{relns}) are labelled by $(\alpha, m)\in R_{N-4},$
\be
\alpha h_m=\sum_{\gamma\in B_{N-4}} \alpha\gamma\,  h_m^\gamma
=\sum_{\beta\in C_{N-4}} M^{(N)}_{\alpha m,\beta}\,\beta,\qquad  \alpha\in C_{N-5}, \;1\leq m\leq N-3,
\ee
where, for $ \alpha\in C_{N-5}, \beta\in C_{N-4},$
\begin{alignat}{3}
M^{(N)}_{\alpha m,\beta}&= h_m^\gamma&&\hbox{if } \beta=\alpha\gamma, \;\gamma\in B_{N-4},\cr
&= 0\qquad&&\hbox{if } \beta\notin\alpha  B_{N-4},\label{MNab}
\end{alignat}
with
\be
\deg M^{(N)}_{\alpha m,\beta}=m+\deg \alpha-\deg\beta,\label{degM}
\ee
and $M^{(N)}_{\alpha m,\beta}=0$ if $\deg\beta-\deg \alpha=\deg \gamma>m$.

The consistency condition is given by the vanishing of
\be\Delta_N=\det M^{(N)}.\ee
 Here, in principle, the $h_m^\gamma$ are evaluated at $x_a=0, 1\leq a\leq N-4,$ but, by the argument used 
 for $N=5,6,7$, the condition $\Delta_N=0$ is independent of the $x_a$.
 
 From (\ref{degM}) it follows that $\Delta_N$ is a homogeneous polynomial of degree $\delta_N$, where
 \begin{align}
\delta_N&= \sum_{m=1}^{N-3} \sum_{\alpha\in C_{N-5}} ( m+\deg \alpha)- \sum_{\beta\in C_{N-4}} \deg\beta\cr
&={(N-3)(N-2)\over 2}|C_{N-5}|+(N-3)d_{N-5}-d_{N-4}
 \end{align}
from which it follows that $\delta_N=(N-3)!,$ the size of $M^{(N)}$. As for $N=6,7,$ since the elements $M^{(N)}_{\alpha m,\beta}$
are at most linear in $v$, and a contribution to the coefficient of $v^{\delta_N}$ can only come from a product of elements of degree one,
$M^{(N)}_{\alpha m,\beta}=h_m^\gamma$, with $\deg\gamma=\deg\beta-\deg \alpha=m-1.$

\subsection{\it Direct demonstration that $\Delta_N$ is the required polynomial.}

Again, write $h=(h_1,h_2,\ldots, h_{N-3})^T$.  First, we can show directly that $\Delta_N=0$ vanishes when $h=0$ because, from (\ref{MNab}), 
in the first column of $M^{(N)}$, 
\be
M^{(N)}_{\alpha m,1}=h_m,\quad\hbox{if}\;\;\alpha=1\;;\qquad M^{(N)}_{\alpha m,1}=0,\quad\hbox{if}\;\;\alpha\ne1,
\ee
and so the whole first column of $M^{(N)}$ vanishes when $h=0$, implying that $\Delta_N=\det M^{(N)}=0$ when $h=0$. 

We have already established that it is a homogeneous polynomial of degree $\delta_N$ and we now wish to show that it is independent of 
$x_b, 1\leq b\leq N-4$. We do this by generalizing the argument of section \ref{directdem7}.
 Let 
\be
C_M^b=\left\{\prod_{a=1\atop a\ne b}^{M} x_a^{m_a}: \;  0\leq m_a\leq M-a+1,\; 1\leq a\leq M, \; a\ne b\right\}
\ee
and
\be
B_M^b=\left\{\prod_{a=1\atop a\ne b}^{M} x_a^{m_a}: \;  0\leq m_a\leq 1,\; 1\leq a\leq M, \; a\ne b\right\}.
\ee
for $1\leq b\leq M$.
Then, the non-zero elements of the column of $M^{(N)}$  specified by 
\be
%:
\beta=x_b^{m_b}\tilde\beta\in C_{N-4},\quad\tilde\beta\in C_{N-4}^b,\quad 0\leq m_b\leq N-b-3\equiv m_b^\ast,
\ee
are in the rows
specified by $(\alpha,m), $ where: $1\leq m\leq N-3; \alpha=x^{m_b'}_b\tilde\alpha,$ and $m_b'=0$ if $m_b=0$, $m_b' =m_b$ or $m'_b=m_b-1$ if $0<m_b<m_b^\ast,$ and $m_b'=m_b^\ast-1$ if $m_b=m_b^\ast$;
and $\tilde \beta =\tilde \alpha\tilde\gamma$ for some $\tilde\gamma\in B_{N-4}$. Then 
\be
 M^{(N)}_{\alpha m,\beta}=h^\gamma_m,\qquad \gamma=x_b^{m_b-m_b'}\tilde\gamma.
\ee
For given $\tilde\beta\in  C_{N-4}^b,$ we replace successively subtract $x_b$ times the column of $M^{(N)}$ labelled by $\beta=x_b^{m_b^\ast-r}\tilde\beta$ from 
that labelled by $\beta=x_b^{m_b^\ast-r-1}\tilde\beta$ for $r=0, 1,\ldots, m_b^\ast-1.$ By this process, the matrix $M^{(N)}$  is transformed into a matrix 
$M^{(N,1)}$ (We suppress the dependence on $b$.) with the non-zero elements in the column labelled by $\beta$ occurring in the rows labelled by 
$(\alpha,m)$  where: $1\leq m\leq N-3;  \alpha=x^{m_b'}_b\tilde\alpha,$ with $\max (0,m_b-1)\leq m_b'\leq m_b^\ast-1$. Then
\begin{alignat}{3}
 M^{(N,1)}_{\alpha m,\beta}&= h^{x_b\tilde\gamma}_m&&\hbox{if } \;m_b'=m_b-1,\cr
&= (-x_b)^{m_b'-m_b}h^{\tilde\gamma}_{m\bu},\qquad&&\hbox{if } m_b-1< m_b'\leq m_b^\ast-1,\label{MN1ab}
\end{alignat}
where $h^{\tilde\gamma}_{m\bu}=h^{\tilde\gamma}_{m}-x_bh^{x_b\tilde\gamma}_{m},$ so that $\partial_{x_b}h^{\tilde\gamma}_{m\bu}=0.$ It follows from the construction 
that $\det M^{(N,1)}=\det M^{(N)}=\Delta_N.$ 

Next add $x_b$ times the row labelled by $(x^{m_b^\ast-r-2}_b\tilde\alpha,m)$ to that labelled by $(x^{m_b^\ast-r-1}_b\tilde\alpha,m)$ for $r=0, 1,\ldots, m_b^\ast-2.$
By this process, the matrix $M^{(N,1)}$  is transformed into a matrix $M^{(N,2)}$ with 
\be
 M^{(N,2)}_{\alpha m,\beta}= M^{(N)}_{\alpha m,\beta}, \quad\hbox{if} \;\;m_b>0;\qquad
 M^{(N,2)}_{\alpha m,\beta}= M^{(N)}_{\alpha m,\beta}-x_b\partial_{x_b}M^{(N)}_{\alpha m,\beta}\quad\hbox{if} \;\;m_b=0,
\ee
and $\det M^{(N,2)}=\det M^{(N,1)}=\Delta_N.$ Now applying these two processes successively to $M^{(N,2)}$, but stopping at $r=m_b^\ast-2$ in the first process and $r=m_b^\ast-3$ 
in the second, we obtain another matrix $M^{(N,4)}$, such that
\be
 M^{(N,4)}_{\alpha m,\beta}= M^{(N)}_{\alpha m,\beta}, \quad\hbox{if} \;\;m_b>1;\qquad
 M^{(N,4)}_{\alpha m,\beta}= M^{(N)}_{\alpha m,\beta}-x_b\partial_{x_b}M^{(N)}_{\alpha m,\beta}\quad\hbox{if} \;\;m_b=0,1,
 \qquad
\ee
$\det M^{(N,4)}=\Delta_N.$ After performing $m_b^\ast$ cycles of this process, we obtain a matrix $M^{(N,2m_b^\ast)}$ all of whose elements are independent of $x_b$ and whose determinant 
equals $\Delta_N$, showing that this also is independent of $x_b$. It follows that $\Delta_N$ has the desired properties of being a homogeneous polynomial of degree $\delta_N$ in $u,v$, 
independent of $x_b, 1\leq b\leq N-4$, vanishing when $h=0$. 

To complete the proof that this polynomial does indeed determine the solutions, we need to show that it does not vanish identically. 
To do this, we show in section \ref{Leading} that only one product in the expression for the determinant $\Delta_N$ contributes to the coefficient of the leading power, $v^{\delta_N}$.

\subsection{The Leading Power of  $v$, $v^{\delta_N}$.}
\label{Leading}

So far we have established that  $\Delta_N=\det M^{(N)}$ is homogeneous of degree $\delta_N=(N-3)!$ in $u,v$, and is independent of $x_a, 1\leq a\leq N-4$.
Thus, to evaluate $\Delta_N$, we can choose eventually to put $x_a=0, 1\leq a\leq N-4$. Because  $h_m^\gamma$ vanishes when  $x_a=0, 1\leq a\leq N-4,$ if
$\deg h_m^\gamma>2$, the only contributions to $\det M^{(N)}$ come from elements
$M^{(N)}_{\alpha m,\beta}=h_m^\gamma$ with $\deg h_m^\gamma=m-\deg\gamma=0,1,2,$ which are constant, linear in $u,v$, and proportional to $uv$, respectively.
 Since each product of $\delta_N$ elements of $M^{(N)}$, contributing to the determinant 
is homogeneous in $u, v$ of degree $\delta_N$, the only products that contribute to the leading term proportional to $v^{\delta_N}$ are ones in which all the factors are of degree 1. These
are of the form  $h_m^\gamma=\partial_\gamma h_m$, where $\deg\gamma=m-1.$ There are 
 $ {N-4\choose m-1}$ such elements $M^{(N)}_{\alpha m,\beta}$ in the row $(\alpha,m)\in R_{N-4}$,
occuring in the columns $\beta=\alpha\gamma,$ where $\deg\gamma=m-1.$ 

To calculate the coefficient of  $v^{\delta_N}$,  let $\widetilde M^{(N)}_{\alpha m,\beta}=M^{(N)}_{\alpha m,\beta}$
if $\deg M^{(N)}_{\alpha m,\beta}=1$, and zero otherwise; then  the coefficient of  $v^{\delta_N}$ in $\det M^{(N)}$ is the same as in $\det \widetilde M^{(N)}$.
Now, $\widetilde M^{(N)}_{\alpha m,\beta}=0$ is zero unless $\deg\beta=\deg\alpha +m-1$, and, if we divide the rows and the columns into subsets
\be
R_{N-4}^j=\{(\alpha,m)\in R_{N-4}:\deg\alpha+m-1=j\},\quad
C_{N-4}^j=\{\beta\in C_{N-4}:\deg\beta=j\},
\ee
$0\leq j\leq d_N=\half(N-3)(N-4),$ the nonzero elements of $\widetilde M^{(N)}$ fall into $d_N+1$ blocks, $\widetilde M_N^j$, with rows labelled by $(\alpha,m)\in R_{N-4}^j$ and columns by $\beta\in C_{N-4}^j,$
$0\leq j\leq d_N.$ These blocks are square with dimension  $d_{Nj}$, where
\be
\prod_{a=1}^{N-4}(1+x+\ldots +x^a)=\sum_{j=0}^{d_N}d_{Nj}x^j=(1-x)^{4-N}\prod_{a=1}^{N-4}(1-x^{a+1}).
\ee
Note that putting $x=1$ in this relation verifies that 
\be
\sum_{j=0}^{d_N}d_{Nj}= (N-3)!=|R_{N-4}|=|C_{N-4}|.
\ee

A non-zero contribution to $\det M^{(N)}$ corresponds to a map $\phi: R_{N-4}\rightarrow C_{N-4}$, which is bijective, such that 
\be\phi(\alpha,m)=\gamma\alpha, \quad\hbox{with}\quad \gamma\in B_{N-4}\quad\hbox{and}\quad  \deg\gamma=m-1.\qquad\qquad[\ast]\ee
 We will show that there is precisely one such map. 
 
 If $m=1$, then $\deg\gamma=0$, so that $\gamma=1$, and hence $\phi(\alpha,1)=\alpha$, for all $\alpha\in C_{N-5}$.

It follows from the block structure of $\widetilde M^{(N)}$ that any such $\phi$ defines bijective maps $R_{N-4}^j\rightarrow C_{N-4}^j,$ and so our task is equivalent to showing that, for
each $0\leq j\leq d_N,$ there is one exactly one bijective map $\phi: R_{N-4}^j\rightarrow C_{N-4}^j$ with the property $[\ast]$. In this case, we shall say the block $\widetilde M_N^j$ has the property $[\ast]$.

We use induction to establish that  $\widetilde M_N^j$ has the property $[\ast]$. 
To this end, we divide $C_{N-4}^{j}$ and $R_{N-4}^{j}$ into subsets $C_{N-4}^{j,k}$ and $R_{N-4}^{j,k}$, respectively. We define
\be
C_n^{j,k}=\left\{\beta\in C_n^{j} :x_{n-k+1}^{k}|\beta\, ; \;  x_a^{n-a+1}\not|\beta, \,1\leq a\leq n-k\right\}
\ee
so that $C_n^{j,k}$ comprises elements $\beta$, which are the product of factors $x_{n-a+1}^{m_a}, 1\leq a \leq n,$ where $0\leq m_a< a$ if $ 1\leq a\leq n-k,$ $m_{n-k+1}=k,$ and
$0\leq m_a\leq a$ if $ n-k+2\leq a\leq n.$ It follows that
\be
C_n^{j,k}=\left\{x_{n-k+1}^{k}\beta': \beta' \in C_{n-1}^{j-k} (x_{1},\ldots,x_{n-k},x_{n-k+2},\ldots,x_{n})\right\}\equiv x_{n-k+1}^{k} C_{n-1}^{j-k}
\ee
For given $N$, $C_{N-4}^{j}$ is the disjoint union of $C_{N-4}^{j}$, $0\leq j\leq d_N=\half(N-3)(N-4);$ for given $N,j,$ $C_{N-4}^{j}$ is the disjoint union of $C_{N-4}^{j,k}$, 
where the range of $k$ is limited above by both $j$ and $N-4$, and below by $0$ and the requirement that $ j-k\leq d_{N-1}$ in order that $C_{N-4}^{j,k}=x_{N-k-3}^{k}C_{N-5}^{j-k}$ be nonempty,
so the range is $\max(0,j-d_{N-1})\leq k\leq \min (j,N-4)$. Note that, in particular, $C_{N-4}^{j,0}=C_{N-5}^{j}.$

Write 
\be
R_{n}^{j,0}=\{(\alpha,1):\alpha\in C_{n-1}, \,\deg\alpha=j\}\subset R_{n}^{j},
\ee
\be
R_{n}^{j,k}=\{(\alpha,m)\in R_{n}^j:1<m\leq n+1\,;\, x_{n-k+1}^{k-1}|\alpha\, ; \;  x_a^{n-a}\not|\alpha, \,1\leq a\leq n-k\}, \qquad 1\leq k\leq j.
\ee

Then $R_{n}^{j,k}=\varnothing $ if $k<j-d_{N-1}$ or $k>N-4$; and $R_{n}^{j}$ is the union of $R_{n}^{j,k}$ for $\max(0,j-d_{N-1})\leq k\leq \min (j,N-4)$. 
Note $|R_{n}^{j,k}|=|C_{n}^{j,k}|$ and
\be
R_{n}^{j,k}=\left\{(x_{n-k+1}^{k-1}\alpha',m'+1): (\alpha',m') \in R_{n-1}^{j-k} (x_{1},\ldots,x_{n-k},x_{n-k+2},\ldots,x_{n-1})\right\},\quad 1\leq k\leq j.
\ee

If $(\alpha,m)\in R_{n}^{j,k}$, with $k> 0$, then $ x_a^{n-a}\not\hskip-3pt|\,\alpha, \,1\leq a\leq n-k$. Suppose $\phi(\alpha,m)=\gamma\alpha$, where $\gamma\in B_n$. Then $x_a^{n-a+1}\not\hskip-3pt|\,\gamma\alpha, \,1\leq a\leq n-k,$ for 
$\gamma$ can add at most one factor of $x_a$ to $\alpha$. Thus $\phi(\alpha,m)\notin C_n^{j,n-a+1}, \, 1\leq a\leq n-k,$ {\it i.e.}
$\phi(\alpha,m)\notin C_n^{j,\ell},\,  k+1\leq \ell\leq n.$ Since $\phi(R_{n}^{j})=C_{n}^{j}$ and $\phi(R_{n}^{j,0})=C_{n}^{j,0}$, because we always have $\phi(\alpha,1)=\alpha$, it follows that
\be
\phi(R_{n}^{j,k})\subset \bigcup_{\ell=1}^k \,C_{n}^{j,\ell}. \qquad 
\ee
In particular $\phi(R_{n}^{j,1})\subset  \,C_{n}^{j,1},$ and since $|R_{n}^{j,1}|=|C_{n}^{j,1}|$ and $\phi$ is bijective, it follows that $\phi(R_{n}^{j,1})=C_{n}^{j,1},$ and similarly,  by induction, that
$\phi(R_{n}^{j,k})=C_{n}^{j,k},$ for all $k$.

Given 
\be
\phi':R_{n-1}^{j-k} (x_{1},\ldots,x_{n-k},x_{n-k+2},\ldots,x_{n-1})\rightarrow C_{n-1}^{j-k} (x_{1},\ldots,x_{n-k},x_{n-k+2},\ldots,x_{n}),
\ee
with the property $[\ast]$, setting $\phi(x_{n-k+1}^{k-1}\alpha',m'+1)=x_{n-k+1}^k\phi'(\alpha',m')$ defines a map 
\be
\phi:R_{n}^{j,k} (x_{1},\ldots,x_{n-1})\rightarrow C_{n}^{j,k} (x_{1},\ldots,x_{n}),
\ee
such that, if $\phi'(\alpha',m')=\gamma'\alpha'$, $\gamma'\in B_{n-1}(x_{1},\ldots,x_{n-1})$, then $\phi(\alpha,m)=\gamma\alpha$, with $\gamma=x_{n-k+1}\gamma'$. So 
$\phi$ has the property $[\ast]$ if $\phi'$ does. Conversely, given $\phi:R_{n}^{j,k} \rightarrow C_{n}^{j,k} $ with the property $[\ast]$, we can construct $\phi':R_{n-1}^{j-k} \rightarrow C_{n-1}^{j-k} $
with the property $[\ast]$, and so the uniqueness of a map $\phi':R_{n-1}^{j-k} \rightarrow C_{n-1}^{j-k} $  with the property $[\ast]$ implies that for a map $\phi:R_{n}^{j,k} \rightarrow C_{n}^{j,k} $
with the property (and conversely). The desired result follows by induction.

Dividing  $C_{N-4}^{j}$ and $R_{N-4}^{j}$ into subsets $C_{N-4}^{j,k}\cong x_{N-k-3}^{k} C_{N-5}^{j-k}$ and $R_{N-4}^{j,k}\cong x_{N-k-3}^{k} R_{N-5}^{j-k}$, respectively, 
and subdividing further and so on inductively defines orders on  $C_{N-4}^{j}$ and $R_{N-4}^{j}$, with respect to which $\phi$ is lower triangular, and, consequently, reordering the rows and columns of $\widetilde M^{(N)}$
brings it into a form which is lower triangular.

It follows that $\det\widetilde M^{(N)}$, which contains the term in $\Delta_N$ involving $v^{\delta_N}$, is given, up to sign, by a single product of the $(N-3)!$ diagonal elements (after reordering as above), which are of the form $h_m^\gamma$, where 
$\gamma\in B_{N-4}$ and $m=\deg\gamma+1$, with each $m$ occurring $(N-4)!$ times. It follows immediately from this that the coefficient of $v^{\delta_N}$ is nonzero provided that all the Mandelstam invariants 
$\sigma_S\ne0$.

To evaluate this coefficient more explicitly, consider further $\det\widetilde M^{(N)}$. Up to sign, it is given by a product of the form
\be
\prod_{\gamma\in B_{N-4}} \left[h^\gamma_{d_\gamma}\right]^{n_\gamma},\quad\hbox{where}\quad d_\gamma=\deg\gamma+1,\label{detwtM}
\ee 
and the sum of the $n_\gamma$ for a given value of $\deg\gamma$ is $(N-4)!$. The number of $\gamma\in B_{N-4}$ with a given value of $\deg\gamma$ is $(N-4)!/ (N-4-\deg\gamma)!(\deg\gamma)!$
and so, if the result is unchanged under permutation of the variables $x_a, 1\leq a\leq N-4$, with respect to which derivatives are taken, $n_\gamma$ will only depend on $\deg\gamma$ and we shall have
\be
    n_\gamma= (N-4-\deg\gamma)!(\deg\gamma)!.\label{ngamma}
    \ee

From this it would follow that the coefficient of $v^{\delta_N}$ is
\be
\prod_{r=0}^{N-4}\prod_{S\subset A''\atop |S|=r}\sigma_{S_{M}}^{n_r},\quad\hbox{where}\quad S_{M}=S\cup\{N-1\},\quad n_r= r!(N-4-r)! 
\ee
where  $A''=\{a:1<a<N-2\}$.

We can establish (\ref{ngamma}) directly from our iterative construction of the map $\phi$. From this construction, if, inductively, $\det\widetilde M^{(N)}$ is given by (\ref{detwtM}), up to a sign,
$\det\widetilde M^{(N+1)}$ is given by 
\be
\left[h_1\right]^{(N-3)!} \prod_{i=1}^{N-3}\prod_{\gamma\in B_{N-4}^i} \left[h^{x_i\gamma}_{d_\gamma+1}\right]^{n_\gamma},\label{detwtM1}
\ee 
where
\be
 B_{N-4}^i=\left\{\prod_{a=1, a\ne i}^{N-3} x_a^{m_a}: \;  0\leq m_a\leq 1,\; 1\leq a\leq N-3\right\}.\label{defBNi}
\ee
Then (\ref{detwtM1}) is exactly what we get from (\ref{detwtM}) and (\ref{ngamma}), replacing $N-4$ by $N-3$, so establishing the result by induction.

\section{Resultants, Hyperdeterminants and Hilbert Series}
\label{Resultant}

\subsection{Sparse Resultants}
\label{Sparse}

It is the special feature of the polynomial scattering equations, 
\be
h_m=0, \qquad 1\leq m\leq N-3,\label{heqs}
\ee
that the $h_m$ are linear and homogeneous in the variables $z_a, 2\leq a\leq N-1,$ and it is this linearity that ensures that we can obtain relatively simple explicit 
forms for the equation $\Delta_N=0,$ obtained by eliminating 
\be
x_a=z_{a+1},\qquad 1\leq a\leq N-4, \label{xa}
\ee
 in favor of $u=z_{N-2}, v=z_{N-1},$ while it is the fact that 
$\deg h_m=m, 1\leq m\leq N-3,$ that implies that the $\deg\Delta_N=\delta_N=(N-3)!$ viewed as a polynomial in $u/v$.

$\Delta_N$ is the resultant of the $N-3$ equations (\ref{heqs}), viewed as polynomial equations in the $N-4$ variables (\ref{xa}), that is 
the polynomial in the coefficients of these polynomials (which are combinations of factors $\sigma_S, u$ and $v$) whose vanishing is the 
necessary and sufficient condition for these equations (\ref{heqs}) to have a common solution for the variables (\ref{xa}), taking values on 
the Riemann sphere. The theory of resultants, 
in the two variable case, has its origins in the work of Sylvester \cite{JJS} and Cayley \cite{AC}, extended to the multivariable case by Macauley 
\cite{Mac1}, and revived and set in a modern context by Jouanolou \cite {Jou} and Gel'fand, Kapranov and Zelevinsky (GKZ) \cite{GKZ}. For accounts, 
see, {\it e.g.,} chapter 4 of \cite{Sturm} or chapter 3 of \cite {DLO}. 

The existence of the resultant, of a set of $M+1$ polynomial equations in $M$ variables, as an irreducible polynomial in the coefficients of the equations with integral 
coefficients, unique up to sign, can be established using techniques of algebraic geometry (see \cite{GKZ} and \cite{Jou}--\cite{DLO}). However, these results do not
provide explicit expressions for the resultant when $M>1$. Moreover, these existence theorems are typically formulated for general polynomials of a particular degree,
involving all its coefficients, and they can vanish identically when one specializes to polynomials in which certain terms are absent. This is the situation for the scattering
equations (\ref{heqs}), where the condition of linearity in each of the variables (\ref{xa}) means that many terms are absent from the general form of a multinomial of 
the same degree. 

Although the theory of resultants for general polynomials can not be directly applied to the scattering equations, the theory of {\it sparse resultants}, developed to deal with situations in which some of the 
monomial terms in the polynomial equations considered are absent, precisely addresses their case (see, {\it e.g.}, section 4.3 of \cite{Sturm}) While a complete theory of sparse resultants is lacking, the scattering equations fall into a category for which some precise and explicit results exist \cite{GKZ}, \cite{GKZ2}--\cite{SZ2}.  In particular, it follows from the discussion of chapter 13 in \cite{GKZ} that the (sparse) resultant of the multilinear equations (\ref{heqs}), viewed as polynomials in the variables (\ref{xa}), defined as above, exists and, by Proposition 2.1 of that chapter, has degree  is $(N-3)!$ (see also \cite{DE}). Because $\Delta_N$, constructed as in
section \ref{construction}, vanishes when (\ref{heqs}) holds, has this degree and has no integer factors, it must indeed be this sparse resultant (which is unique up to sign), and so irreducible as a polynomial.
Consequently, there is a solution to (\ref{heqs}) for  (\ref{xa}) for each of the $\delta_N$ roots of $\Delta_N=0$ for $u/v$.

\subsection{Hyperdeterminants and the Scattering Equations}

Gel'fand, Kapranov and Zelevinsky \cite{GKZ} relate their discussion of sparse resultants to a theory of multidimensional or hyperdeterminants, which they have developed in this context. This theory generalizes the familiar concept of the determinant of a (two-dimensional) square matrix, $\M_{i_0i_1}, 0\leq i_0,i_1\leq\ell,$  of shape $(\ell+1)\times (\ell+1)$, to an $(M+1)$-dimensional matrix or array,
$\M_{i_0i_1\cdots i_M}$, where $0\leq i_j\leq\ell_j, 0\leq j\leq M,$  of shape or format $(\ell_0+1)\times (\ell_1+1)\times \cdots\times (\ell_M+1),$ and we take $\ell_0$ to be the  largest of the $\ell_j,$ that is
$\ell_j\leq\ell_0, 1\leq j\leq M.$ Associated to this multidimensional matrix are the multilinear functions
\be
f_{i_0}(\xi_1,\ldots,\xi_M)=\M_{i_0i_1\cdots i_M}\xi_1^{i_1}\ldots\xi_M^{i_M}, \qquad 0\leq i_0\leq \ell_0,\label{deff}
\ee
where implicitly $i_j$ is summed over $0\leq i_j\leq\ell_j,$ and $\xi_j\in\Cop^{\ell_j}, 1\leq j\leq M.$ 

The hyperdeterminant, $\det\M,$ is defined as the (sparse) resultant of the $\ell_0+1$ homogeneous equations 
\be
f_{i_0}(\xi_1,\ldots,\xi_M)=0, \qquad 0\leq i_0\leq \ell_0,\label{feqs}
\ee
that is the polynomial in the coefficients of the $f_{i_0}$ (the elements of the  multidimensional matrix $\M$) 
whose vanishing is the necessary and sufficient condition for (\ref{feqs}) to have a nontrivial solution for the $\xi_j$,
which we should regard as elements of $\Cop\Pop^{\ell_j}$, because of the multihomogeneity of (\ref{deff}). 
$\det\M$ itself is nontrivial if and only if $\ell_0\leq \ell_1+\cdots+\ell_M,$ for otherwise the $\xi_j\in\Cop\Pop^{\ell_j}$
are always overdetermined. This definition of $\det\M$ agrees with the conventional definition in the two-dimensional 
case.

Explicit formulae for $\det\M$ in the general case are not known, but an exception is the borderline case, when $\ell_0= \ell_1+\cdots+\ell_M;$ 
in this case $\M$ is said to be of {\it boundary format},  there is a simple expression for the degree, $\deg\M=(\ell_0+1)!/\ell_1!\cdots \ell_M!,$
and $\det\M$ can constructed as a (two-dimensional) determinant, using methods similar to those of  sections \ref{Elimination7}
and \ref{EliminationG}.

$\Delta_N$ is the hyperdeterminant of a multidimensional matrix of boundary format, $(N-3)\times 2\times \cdots\times 2$, where there are $M=N-4$ factors of $2$. 
To show this, we homogenize $h_m$ as a function of the variables (\ref{xa}) by setting $x_a=\xi_a^1/\xi_a^0$, and defining 
\be
f_{m-1}(\xi_1,\ldots,\xi_M)=h_m\left(\xi_1^1/\xi_1^0,\ldots,\xi_{M}^1/\xi_{M}^0,u,v\right)\prod_{a=1}^M\xi_{a}^0,\qquad 1\leq m\leq M+1=N-3,
\label{deffh}
\ee
so that the $f_m$ are linear in each $\xi_a$. Then, writing $A^\vee=\{a:1\leq a\leq N-4\},$
\begin{align}
f_{m-1}(\xi_1,\ldots,\xi_M)&=\sum_{U\subset A^\vee\atop |U|=m} \sigma_{\widetilde U}\,\xi_U^1\,\xi_{\overline U}^0\,+\sum_{U\subset A^\vee\atop |U|=m-1} \sigma_{\widetilde U_u}\,\xi_U^1\,\xi_{\overline U}^0\,u\cr
&\qquad+\sum_{U\subset A^\vee\atop |U|=m-1} \,\sigma_{\widetilde U_v}\,\xi_U^1\,\xi_{\overline U}^0\,v+\sum_{U\subset A^\vee\atop |U|=m-2} \sigma_{\widetilde U_{uv}}\,\xi_U^1\,\xi_{\overline U}^0\,u\,v
\end{align}
where $\overline U$ denotes the complement of $U$ in $A^\vee$, $\widetilde U=\{a+1: a\in U\}, \widetilde U_u=\widetilde U\cup\{N-2\},\widetilde U_v=\widetilde U\cup\{N-1\},$ and $ \widetilde U_{uv}=\widetilde U\cup\{N-2,N-1\}.$ This determines a multidimensional matrix $\M$ as in (\ref{deff}) with $\ell_0=M$ and $\ell_a=2, 1\leq a \leq M$.
Since the condition that the equations $h_m=0$ have solutions for some $x_a$ clearly the same as that the equations $f_{m-1}=0$ have solutions for some $\xi_a$, it follows that $\Delta_N=\det\M$, up to sign.

We note that the hyperdeterminant corresponding to $N=6$ case is discussed by GKZ as example 4.9 of \cite{GKZ2};
and, as Cardona and Kalousios \cite{CK} have pointed out, Sturmfels discussed the equations of the $N=7$ case as an example
of sparse resultants in section 4.5 of \cite{Sturm}.

\subsection{Hilbert Series and Regular Sequences}

The CHY expressions for tree amplitudes in massless field theories  are given as residue functions evaluated at the solutions of the 
scattering equations and summed over those solutions. Thus  such a residue function, $\Phi$,  only need be specified up the 
addition of sums of polynomial multiples of the $h_m,$ because the additional terms vanish at the solutions of the scattering equations.

If $\R$ denotes the ring of polynomials in $z_a, 2\leq a\leq N-1$, and $\I\equiv\langle h_1,\ldots ,h_{N-3}\rangle$ denotes the ideal in $\R$ generated by the $h_m, 1\leq m\leq N-3,$
that is the polynomials of the form $g_1h_1+\ldots+g_{N-3}h_{N-3}$, we are interested in $\S=\R/\I$, polynomials modulo elements of $\I$. Further, we 
consider homogeneous polynomials in the $z_a$ of a specific degree, $n$, say. Suppose $\R_n$ denotes the homogeneous polynomials of
degree $n$, so that $\R$ is the direct sum of the $\R_n$; and $\I_n$ denotes the elements of $\I$ which are homogeneous of degree $n$, so that 
$\I_n=\I\cap\R_n$. Then if $\S_n=\R_n/\I_n$, $\S$ is the direct sum of the $\S_n$ and $H^\S_n=\dim\S_n$, the number of independent homogeneous
functions of degree $n$ modulo the $h_m$, which is the number of independent functions we have to consider. These dimensions are encoded into a formal series, the Hilbert series (see, {\it e.g.,} \cite{Eis}), 
\be
H^\S(t)=\sum H^\S_nt^n.
\ee

If $\J$ is a homogeneous ideal in $\R$, $\T=\R/\J$ and $h\in\R_s$, a homogeneous polynomial of degree $s$, $h\notin\J,$ we can consider the 
ideal $\K,$ generated by $\J$ and $h$, {\it i.e.} $\K=\{g_1+hg_2:g_1\in\J,g_2\in\R\}$, and the corresponding quotient ring, $\U=\R/\K$. Define 
$\phi:\T\rightarrow\T$ by $\phi(g+\J)=gh+\J.$ The kernel of $\phi$ consists of those $(g+\J)\in\T$, for which $gh\in\J$, so $\ker\phi$ is trivial if
$gh\notin\J$ whenever $g\notin\J$. If there is $g\notin\J$ with $gh\in\J$, $h$ is said to be a zero divisor in $\T$. If $h$ is  a nonzero divisor in $\T,$
$\ker\phi$ is trivial, and it is easy to calculate the Hilbert series of $\U$ in terms of that of $\T$.

Because $\J\subset\K$, we can define $\psi:\T\rightarrow\U$ by $\psi(g+\J)=g+\K$. The kernel of $\psi$ consists of $(g+\J)\in\T$ such that $g\in\K$,
implying $g+\J=g'h+\J$, for some $g'$, implying $\ker\psi=\hbox{im}\,\phi,$ while, clearly, $\hbox{im}\,\psi=\U$.
If $\deg h=d$, we have maps $\phi:\T_n\rightarrow\T_{n+d}$ and 
$\psi:\T_n\rightarrow\U_{n}$. So, provided that $h$ is  a nonzero divisor in $\T,$ we have an exact sequence:
 \begin{center}
\begin{tikzpicture}[>=angle 90,scale=2.2,text height=1.5ex, text depth=0.25ex]
%%First place the nodes
\node (a-1) at (0,0) {};
\node (a0) [right=of a-1] {$0$};
\node (a1) [right=of a0] {$ \T_n$};
\node (a2) [right=of a1] {$\T_{n+d}$};
\node (a3) [right=of a2] {$\U_{n+d}$};
\node (a4) [right=of a3] {$0$};
%%Draw the black arrows
\draw[->,font=\scriptsize]
(a0) edge (a1)
(a2) edge node[auto] {$\psi$} (a3)
(a3) edge (a4)
(a1) edge node[auto] {$\phi$} (a2);
\end{tikzpicture}
\end{center}
$\dim\T_n=\dim[\im\phi]_{n+d},$ $\dim\T_{n+d}=\dim[\ker\psi]_{n+d}+\dim[\im\psi]_{n+d}=\dim\T_n+\dim\U_{n+d},$ using
 $\im\phi=\ker\psi$.
Multiplying by $t^{n+d}$ and summing gives
\be
H^\U(t)=(1-t^{d})H^\T(t),\label{Hut}
\ee
provided that $h$ is  a  non-zero divisor in $\T$.

We can use this result to calculate $H^\S(t)$ iteratively if $h_m$ is a nonzero divisor in $\R/\I^{(m-1)}$, $2\leq m\leq N-3$, where
$\I^{(m)}$ denotes the ideal $\langle h_1,\ldots ,h_{m}\rangle$, generated by $h_1,\ldots ,h_{m}$. In this case, $h_1,\ldots ,h_{N-3}$ is said to be a {\it regular sequence},
and we can apply (\ref{Hut}) iteratively to obtain 
\be
H^{\S^{(m)}}(t)=(1-t^{m})H^{\S^{(m-1)}}(t),\qquad 1\leq m\leq N-3,
\ee
where $\S^{(m)}=\R/\I^{(m)},\, \S^{(0)}=\R$. So,   since $H^{\R}(t)=(1-t)^{2-N},$
\be
H^{\S}(t)={1\over (1-t)^{N-2}}\prod_{m=1}^{N-3}(1-t^m)={1\over 1-t}\prod_{m=2}^{N-3}(1+t+\ldots+t^{m-1}),\label{HSeries}
\ee
consistent with results obtained by computer calculation in \cite{HMS} for low values of $N$.

That, in the case of the scattering equations,  $h_1,\ldots ,h_{N-3}$ is indeed a regular sequence follows from the {\it Unmixedness Theorem} of Macaulay \cite{Mac2}. (For an
exposition in modern terminology, see, {\it e.g.,} \cite{MSa}, specially section 5.) This theorem states that, if $g_1, g_2, \ldots g_s$ is a regular sequence of homogeneous polynomials, 
in the $M$ (complex) variables, $x_1,x_2,\ldots,x_M,$ then the (projective) variety defined  by $g_1=g_2=\cdots=g_s=0$ has dimension $M-s$ ({\it i.e.} projective dimension $M-s-1$), and, 
conversely, if this variety has dimension $M-s$ then $g_1, g_2, \ldots g_s$ is a regular sequence (irrespective of the order). Taking $M=N-2$, with $x_a=z_{a+1}, 1\leq a\leq N-2,$ we have established in section
\ref{EliminationG} that the variety defined by the scattering equations (\ref{heqs}) is discrete, {\it i.e.} has projective dimension zero, so that, by the Unmixedness Theorem, $h_1, h_2, \ldots,
h_{N-3}$ is a regular sequence.

In fact, we do not have rely on the analysis of section \ref{EliminationG} to show that we have a regular sequence. If we supplement the $N-3$ scattering equations (\ref{heqs}) with the equation $z_{N-1}=0$, 
there is no solution to this augmented system, other than the trivial solution in which all the $z_a$ vanish, because we have already specialized to $z_N=0$, and we established in \cite{DG2}  that the values of the 
$z_a$ must be distinct for different $a$ if the Mandelstam variables $\sigma_S=k^2_{S_1}$ are nonzero. [We can also deduce this directly by first substituting $z_{N-1}=0$ in $h_{N-3}=0$ to obtain 
\be
\sigma_{A''}\,z_2z_3\cdots z_{N-2}=0,\label{sz}
\ee
where $A''=\{a\in A: a\ne 1, N-1,N\}$, implying that one of $z_a, 2\leq a\leq N-2,$ vanishes, $z_b$ say. Then, putting $z_b=z_{N-1}=0$ in $h_{N-2}=0$, we can deduce that another $z_a$ vanishes and so on to
establish that all the $z_a, 2\leq a\leq N-1,$ vanish, showing again that there is no nontrivial solution to (\ref{heqs}) together with $z_{N-1}=0$.] So the space of solutions has (affine) dimension zero, {\it i.e.} is empty projectively, and by the Unmixedness Theorem, $h_1, h_2, \ldots, h_{N-3}, z_{N-1}$ is a regular sequence, and it follows that the subsequence $h_1, h_2, \ldots, h_{N-3}$ is regular. 

[A somewhat similar argument has been used in \cite{BSZ} to show that the (nonhomogeneous) polynomials obtained by putting $z_2=1$ in $h_1,\ldots ,h_{N-3}$ form a H-basis, in the sense of Macauley, for the (affine) ideal that they generate.]

The regularity of the sequence of scattering equations implies that the formula (\ref{HSeries}) for the Hilbert series holds.  $H^\S_n$, as a function of $n$, is called the Hilbert function. For sufficiently large $n$, it is always a polynomial whose degree equals the projective dimension of the variety, and so is eventually constant for a zero-dimensional variety, with the constant equalling the number of points in the discrete variety, counted according to multiplicity. In this case, $H^\S_n = (N-3)!$ for $n\geq \half (N-3)(N-4)$, corresponding to the fact that the scattering equations have $(N-3)!$ solutions.

The techniques of this subsection can be applied to the generalizations of the scattering equations, corresponding to M\"obius spin $\half N -\ell$, $2\leq \ell\leq \half N$, introduced in \cite{DG2}. (The values $\ell =0,1$ correspond to overdetermined systems.) These are systems of 
$N-2\ell+1$ homogeneous equations, in $N-2$ variables, $z_a$, of each degree from $\ell -1$ to $N-\ell-1$. ($\ell=2$ corresponds to polynomial form of the original CHY scattering equations.) These equation are linear in each of the $z_a$. Provided that the coefficients in the polynomials are all nonzero, after supplementing the $N-2\ell+1$ homogeneous polynomials by $2\ell -1$ of the $z_a$, we can use a argument as in (\ref{sz}) to show that this system of $N-2$ equations in $N-2$ variables has only a trivial solution. So, again, the polynomials form a regular sequence and the Hilbert series for the associated ring, $\S=\R/\I$, where $\I$ is the ideal
generated by the  $N-2\ell+1$ homogeneous polynomials is given by 
\be
H^{\S}(t)={1\over (1-t)^{N-2}}\prod_{m=\ell-1}^{N-\ell-1}(1-t^m)={1\over (1-t)^{2\ell-3}}\prod_{m=\ell-1}^{N-\ell-1}(1+t+\ldots+t^{m-1}),\label{HSeries2}
\ee
  so that the $H^\S_n$ is a polynomial of degree $2\ell-4$ for sufficiently large $n$, corresponding to a variety of this projective dimension. This expression for $ H^{\S}(t)$ is again consistent with results  in \cite{HMS} for low values of $N$.

\section*{Acknowledgements}
LD was partially supported by NSF grant No. PHY-1620311, 
and thanks the Institute for Advanced Study at Princeton for its hospitality.
PG was partially supported by NSF
grant No. PHY-1314311.

\vskip20pt
\singlespacing

%%% References %%%

\providecommand{\bysame}{\leavevmode\hbox to3em{\hrulefill}\thinspace}
\providecommand{\MR}{\relax\ifhmode\unskip\space\fi MR }
\providecommand{\MRhref}[2]
{%\href{http://www.ams.org/mathscinet-getitem?mr=#1}{#2}
}
\providecommand{\href}[2]{#2}

\end{document}